\documentclass[journal,twoside,web]{file/ieeecolor}
\usepackage{file/generic}
\usepackage{cite}
\usepackage{amsmath,amssymb,amsfonts}
\usepackage{algorithmic}
\usepackage{textcomp}
\usepackage{mathrsfs}
\usepackage{graphicx}
\graphicspath{{file/}{figure/}}

\newtheorem{remark}{\bf Remark}
\newtheorem{assumption}{\bf Assumption}
\newtheorem{lemma}{\bf Lemma}
\newtheorem{definition}{\bf Definition}
\newtheorem{corollary}{\bf Corollary}
\newtheorem{theorem}{\bf Theorem}

\newcommand{\refeq}[1]{(\ref{#1})}

\begin{document}
\title{Structure Identification of NDS with Descriptor Subsystems under Irregular Sampling}

\author{Yunxiang Ma, Tong Zhou, \IEEEmembership{Fellow, IEEE}
	\thanks{This work was supported in part by the NNSFC under Grant 62127809, 62373212, 61733008, and 52061635102.}
	\thanks{Yunxiang Ma is with the Department of Automation, Tsinghua University, Beijing 100084, China. Email: {\tt\small mayx23@mails.tsinghua.edu.cn}}
	\thanks{Tong Zhou is with the Department of Automation and BNRist, Tsinghua University, Beijing, 100084, China. Email: {\tt\small tzhou@mail.tsinghua.edu.cn}}
}

\maketitle

\begin{abstract}
This paper extends previous identification method to the asynchronous sampling scenario, enabling the simultaneous handling of asynchronous, non-uniform, and slow-rate sampling conditions. Moving beyond lumped systems, the proposed framework targets the identification of interconnection structure of Networked Dynamic Systems (NDS) with descriptor-form subsystems. In the first stage, right tangential interpolations are estimated from steady-state outputs, allowing all asynchronous samples to be fused into a unified estimator. In the second stage, a left null-space projection is employed to decouple the bilinear dependence between state-related matrices and interconnection parameters, reducing the identification problem to two successive linear estimation problems. The proposed approach eliminates the full-normal-rank transfer matrix assumption required in previous work, while providing theoretical guarantees of mean-square consistency and asymptotic unbiasedness. Numerical results demonstrate that the framework can accurately recover the system structure, even under severe sampling irregularities.

\end{abstract}

\begin{IEEEkeywords}
	Structure identification, Tangential interpolation, Irregular sampling, Networked dynamic system
\end{IEEEkeywords}


\section{Introduction}

\subsection{Motivation and Practical Significance}

Networked Dynamic Systems (NDS), which are composed of many interconnected subsystems, are ubiquitous in areas ranging from electrical power grids to complex biological networks. The overall behavior and performance of an NDS depends not only on the internal dynamics of each subsystem, but more importantly on the network topology that defines their interconnections \cite{Chetty2020}. Consequently, accurately identifying the interconnection structure from measured input-output data is crucial for system analysis and control, enabling controller synthesis, fault diagnosis, and the assurance of operational reliability \cite{Weerts2018}.

However, in practical applications, a major challenge is that the data we collect are often irregular. Subramanian et al. \cite{Subramanian2020} show that when corrupt measurement data are used to identify the structure of linear time-invariant (LTI) systems, the process can result in the inference of spurious or erroneous links in the network. Unlike laboratory settings with carefully controlled conditions, real-world systems often involve geographically dispersed devices, constrained communication, sensor failure and event-driven operation \cite{Feichtinger2021}. These factors lead to several types of sampling irregularities:

\begin{itemize}
	\item \textbf{Asynchronous Sampling:} Subsystems may operate with independent clocks or be separated by long distances, making precise temporal alignment of measurements  across the entire network infeasible.
	\item \textbf{Non-Uniform Sampling:} Sensor faults, packet losses, or temporary communication delays can result in missing data or variations in sampling intervals. This variability can degrade the performance of methods that assume uniform sampling.
	\item \textbf{Low-Rate Sampling:} Some state variables are costly or energy-intensive to measure, or they are sampled by event-driven sampling strategies. As a result, the resulting sampling rates may fall below the Nyquist frequency.
\end{itemize}

These three types of sampling irregularity frequently occur together, posing a significant challenge to conventional system identification methods, which typically assume that data are uniform, synchronized, and sampled at a sufficiently high rate.

\subsection{Literature Review and Existing Gaps}

While substantial research has addressed dynamic system identification, most existing methods struggle with the sampling irregularities commonly encountered in real-world networked systems. Current approaches typically handle only one or two types of irregular sampling conditions simultaneously, with no existing method capable of addressing all three concurrently.

Regarding asynchronous sampling, this challenge is particularly prevalent in networked systems where subsystems operate independently. Early work by Zhang et al. \cite{Zhang2001} analyzed the stability implications of asynchronous sampling in networked control systems. More recently, Li et al. \cite{Li2023a} studied asynchronous sampling in LTI systems, focusing specifically on secure state estimation. They proposed a decentralized local estimation algorithm that ensures secure estimates even under attacks. In the context of distributed estimation, Battistelli and Chisci \cite{Battistelli2014} developed consensus-based approaches for handling asynchronous measurements in multi-sensor networks, providing theoretical guarantees for stability and convergence.

For non-uniform sampling, Heemels et al. \cite{Heemels2010} provided a comprehensive framework for understanding the tradeoffs between transmission intervals, delays, and system performance under communication constraints. Lin and Sun \cite{Lin2019} developed an optimal distributed fusion estimation algorithm for systems with non-uniform sampling, correlated noises, and packet dropouts. Additionally, Cheng et al. \cite{Cheng2024} introduced a sojourn probability strategy to manage non-uniform sampling in fuzzy sampled-data systems, effectively capturing the random behavior of sampling periods while ensuring mean-square stability. Hu et al. \cite{hu2024} proposed an approach utilizing a lifted state-space model combined with eigenvalue decomposition to identify multirate system dynamics, effectively handling non-uniform data through collaborative sensing mechanisms.

Concerning slow-rate sampling, Yue et al. \cite{Yue2020} presented a heuristic algorithm that imposes sparsity constraints and handles matrix functions to reconstruct sparse networks from full-state measurements with low sampling frequencies, assuming the absence of aliasing effects. Huang et al. \cite{Huang2021} introduced an identification method based on invariant-subspace theory, combining the advantages of both time-domain and frequency-domain approaches. This method was subsequently extended in \cite{Huang2023} to simultaneously handle non-uniform and slow sampling conditions, with the capability to accommodate asynchronous input and output channels. However, its applicability was limited to single-input single-output (SISO) systems with periodic input signals.

More recently, Zhou \cite{Zhou2024a} investigated the time-domain identification of continuous-time multi-input multi-output (MIMO) descriptor systems under slow and non-uniform sampling conditions, providing explicit formulas for both transient and steady-state responses along with a parametric estimation algorithm. However, this method's applicability was constrained by a restrictive algebraic condition: it required either the transfer function matrix $G_{yv}(s)$ (from internal inputs to external outputs) or $G_{zu}(s)$ (from external inputs to internal outputs) to have full normal column or row rank, respectively. This condition is frequently violated in practice, particularly in systems with specific physical constraints or structural properties, thereby significantly limiting the method's scope of application.

The concept of tangential interpolation, while well-established in model reduction theory, has seen limited application in system identification from time-domain data. Tangential interpolation represents the directional evaluation of matrix-valued functions at specific complex points. Recent work by Astolfi \cite{Astolfi2010} and Zhou \cite{Zhou2024a} revealed the intrinsic connection between tangential interpolations of transfer functions and steady-state responses of LTI systems under specific input stimulation. This insight has sparked development of data-driven approaches: the Loewner framework by Mayo and Antoulas \cite{Mayo2007}, extended by Ionita and Antoulas \cite{Ionita2014}, provides systematic data-driven model reduction through tangential interpolation. Bhattacharjee and Astolfi \cite{Bhattacharjee2024} further extended this to systems with unknown inputs, enabling direct reduced-order model generation from data.

\subsection{Main Contributions}

This paper addresses the aforementioned gaps by proposing a systematic approach for identifying the interconnection structure of continuous-time MIMO NDS, using time-domain data collected under asynchronous, non-uniform, and slow sampling conditions.

We extend previous two-stage identification algorithm \cite{Zhou2024a} to the asynchronous sampling setting, enabling the simultaneous handling of all three types of irregularity. In the first stage, the right tangential interpolations of the overall NDS transfer function are estimated from the steady-state outputs. These interpolations provide a nonparametric characterization that naturally integrates asynchronous, non-uniform, and slow-sampled data into a unified estimator. In the second stage, the estimated interpolations are used to recover the unknown interconnection parameters.

The core challenge in this problem lies in the bilinear coupling between the state-dependent matrices and the interconnection parameters $\theta$. We resolve this issue via an algebraic decoupling based on a left null-space projection, which eliminates all $\theta$-dependent terms and leads to a linear system for an intermediate state-related matrix. Once this matrix is uniquely identified, the interconnection parameters can be determined through a standard linear least-squares procedure.

This paper removes the restrictive full-rank requirements on transfer matrices imposed in prior work \cite{Zhou2024a}, thereby broadening the class of identifiable networked systems. While \cite{Zhou2024a} handles non-uniform and slow sampling for a single lumped system, this paper extends it to handle asynchronous sampling across a distributed network. It allows to unify sparse, uncoordinated local measurements into one consistent global estimation problem. Unlike the black-box identification method in \cite{Huang2023}, our approach explicitly identifies the physical interconnection structure without requiring any specific type of excitation input.

In summary, the main contributions of this paper are as follows:

\begin{itemize}
	
	\item \textbf{Unified framework for irregular sampling:} We extend previous two-stage identification algorithm \cite{Zhou2024a} capable of simultaneously handling asynchronous, non-uniform, and slow sampling scenarios. The method introduces a fusion mechanism based on the right tangential interpolations of the system transfer function, transforming each isolated measurement from any subsystem into a linear constraint on a shared global interpolation vector.
	
	\item \textbf{Expanded applicability through bilinear decoupling:} A left null-space projection is employed to remove the bilinear dependence between state-related matrices and the interconnection parameters, resulting in two consecutive linear estimation steps. This decoupling eliminates the restrictive full-rank conditions required by our earlier work \cite{Zhou2024a} and substantially expands the range of identifiable networked systems.
	
\end{itemize}

The effectiveness of the proposed method is validated through numerical simulations, demonstrating its ability to accurately recover interconnection parameters under irregular sampling conditions and its robustness against the local minima problem that plagues traditional nonlinear optimization methods.

\subsection{Paper Organization}

The remainder of this paper is organized as follows. Section II presents the problem formulation, detailing the NDS model, the input signal generation, and the key assumptions. Section III develops the presents the two-stage structure identification algorithm, detailing the estimation of the right tangential interpolations and subsequent parameter reconstruction. Section IV provides an analysis of the algorithm's properties and Section V validates the algorithm's effectiveness through numerical simulations. Section VI concludes the paper.

The following notation will be employed throughout this paper. The superscripts $\star^*$ and $\star^{\dagger}$ denote the conjugate transpose and Moore-Penrose pseudoinverse, respectively. The operator $\operatorname{vec}\{\star\}$ represents the vectorization of a matrix, while $\operatorname{diag}\left\{\left.\star_i\right|_{i=1}^n\right\}$, $\operatorname{col}\left\{\left.\star_i\right|_{i=1}^n\right\}$, and $\operatorname{row}\left\{\left.\star_i\right|_{i=1}^n\right\}$ construct matrices by diagonally, vertically, and horizontally stacking the elements $\left.\star_i\right|_{i=1}^n$, respectively. For a complex vector or matrix, $\operatorname{Re}\{\cdot\}$ and $\operatorname{Im}\{\cdot\}$ denote the operations of taking its real and imaginary parts, respectively.

\section{Problem Formulation and Preliminaries}

This section introduces the NDS system model adopted in this paper and the model for generating input signals. We then formally state the identification problem and present the key assumptions and preliminary results that form the foundation of our subsequent analysis.

\subsection{Model of the NDS}

For a NDS composed of $N$ interconnected subsystems, the dynamics of the $i$-th subsystem $\Sigma_i$ can be described in descriptor form as below, which is believed to be more natural and convenient for expressing constraints among system variables and preserving the systems’ structural information \cite{Duan2010}.
\begin{equation}
	\label{eq:subsystem}
	\begin{bmatrix} {E}_{i}{\dot{x}}_{i}(t) \\ {z}_{i}(t) \\ {y}_{i}(t) \end{bmatrix} = \begin{bmatrix} {A}_{xx}^{\left[ i\right]} & {B}_{xv}^{\left[ i\right]} & {B}_{xu}^{\left[ i\right]} \\ {C}_{zx}^{\left[ i\right]} & {D}_{zv}^{\left[ i\right]} & {D}_{zu}^{\left[ i\right]} \\ {C}_{yx}^{\left[ i\right]} & {D}_{yv}^{\left[ i\right]} & {D}_{yu}^{\left[ i\right]} \end{bmatrix} \begin{bmatrix} {x}_{i}(t) \\ {v}_{i}(t) \\ {u}_{i}(t) \end{bmatrix} + \begin{bmatrix} 0 \\ 0 \\ {n}_{i}(t) \end{bmatrix}
\end{equation}

Here, for the $i$-th subsystem $\Sigma_i$, ${x}_{i}(t)$ is the state vector, and the real matrix ${E}_{i}$ may be singular. The vectors ${u}_{i}(t)$ and ${y}_{i}(t)$ are the external input and output, respectively. The internal input ${v}_{i}(t)$ and internal output ${z}_{i}(t)$ represent signals exchanged with other subsystems. The term $n_i(t)$ represents the noise affecting the external output vector $y_i(t)$, which is assumed to be uncorrelated at each sampling instant.

The interconnections among the subsystems of the whole NDS $\Sigma_\theta$ are described by the following equation:
\begin{equation}
	\label{eq:interconnection}
	v(t)=\Phi(\theta) z(t)
\end{equation}
where $v(t)=\operatorname{col}\left\{\left.v_i(t)\right|_{i = 1}^N\right\}$ and $z(t)=\operatorname{col}\left\{\left.z_i(t)\right|_{i = 1}^N\right\}$ are the stacked internal input and output vectors for the entire NDS. Similarly, $x(t)$, $y(t)$, $z(t)$ and $n(t)$ can be defined. The matrix $\Phi(\theta)$, which describes the interconnections among subsystems, is called the topology or structure of the overall NDS $\Sigma_\theta$. Here, $\Phi(\theta)$ is assumed to be affine with respect to $\theta=\left(\theta_1, \theta_2, \cdots, \theta_{m_\theta}\right)^T \in \Theta$, i.e.,
\begin{equation}
	\label{eq_Phi_theta}
	\Phi(\theta) = \Phi_0 + \sum_{i = 1}^{m_{\theta}} \theta_i \Phi_i
\end{equation}
where $m_{\theta}$ is the number of unknown parameters in the model. For $i = 0, 1, 2, \cdots, m_{\theta}$, $\Phi_i$ are known real matrices that represent prior structural information about the system structure from its working principles. If no prior information is available and the objective is to identify whether any interconnections exist, the entire matrix $\Phi$ can be regarded as an unknown variable to be estimated. The parameters $\theta_i$ are independent unknown parameters, and the set $\Theta \subseteq \mathbb{R}^{m_{\theta}}$ specifies their admissible range.

In this paper, the dimension of the vector $\star_i(t)$ is denoted by $m_{\star, i}$, where $i = 1,2, \ldots, N$ and $\star \in \{x,y,z,u,v\}$. Using these notations, we define the integer $m_{\star}=\sum_{i = 1}^N m_{\star, i}$. Then, $\Phi(\theta)$ is a real matrix of dimension $m_v \times m_z$.

Let $E=\mathrm{diag}\left\{ \left. E_i \right|_{i=1}^{N} \right\}$, $X _{\star \sharp}=\mathrm{diag}\left\{ \left. X _{\star \sharp}^{[i]} \right|_{i=1}^{N} \right\}$, where $X \in \{A,B,C,D\}$, $\star \in \{y, z\}$ and $\sharp \in \{u, v\}$. 


Based on \eqref{eq:subsystem} and \eqref{eq:interconnection}, the overall dynamics of the NDS $\Sigma_{\theta}$ can also be expressed as:
\begin{equation}
	\begin{aligned}
		&\bar{E}\begin{bmatrix}
			\dot{x}(t) \\
			\dot{z}(t)
		\end{bmatrix}=A_{\theta}\begin{bmatrix}
			x(t) \\
			z(t)
		\end{bmatrix}+\begin{bmatrix}
			B_{x u} \\
			D_{z u}
		\end{bmatrix} u(t)\\
		&y(t)=C_\theta \begin{bmatrix}
			x(t) \\
			z(t)
		\end{bmatrix}+D_{y u} u(t)+n(t)\\
	\end{aligned}	
\end{equation}
where $\bar{E}= \begin{bmatrix}
	E&		0\\
	0&		0\\
\end{bmatrix} $, $A_{\theta} = 
\begin{bmatrix}
	A_{xx} & B_{xv}\Phi(\theta) \\
	C_{zx} & D_{zv}\Phi(\theta) - I_{m_z} \\
\end{bmatrix} $ and $C_\theta = \begin{bmatrix}
	C_{y x} & D_{y v} \Phi(\theta)
\end{bmatrix}$. 

For the NDS to be physically meaningful and analyzable, we impose the following standard assumptions.
\begin{assumption}
	\label{assum_1}
	For any admissible parameter vector $\theta \in \Theta$, all subsystems $\left.\Sigma_i\right|_{i = 1}^N$ are regular, and the entire NDS $\Sigma_\theta$ is well-posed and regular. In addition,the set $\mathbf{\Theta }$ is compact, meaning it is bounded and closed.
\end{assumption}

Assumption \ref{assum_1} is the basic requirement for the normal operation of the system \cite{Zhou2020}, and it is also the prerequisite for structure identification. It indicates that the time-domain model of the entire NDS is well-defined, ensuring that the system responds in a deterministic manner to admissible external inputs and initial conditions.

\subsection{Model of Input-Generation System}

The external input signal $u(t)$ of the NDS $\Sigma_\theta$ is generated by the following LTI autonomous system $\Sigma_s$ with $m_\xi$ states and $m_u$ outputs:
\begin{equation}
	\label{eq_signal_generation}
	\begin{aligned}
		&\dot{\xi}(t) = \Xi \xi(t)\\
		&u(t) = \Pi \xi(t)
	\end{aligned}
\end{equation}
where $\xi(t) \in \mathbb{R}^{m_\xi}$ is the state of the generator. The state matrix $\Xi $ being real implies its eigenvalues are either real or appear in complex conjugate pairs, and corresponding eigenvectors have the same relationship. Let $ \lambda_{r,i} $ ($ i = 1, 2, \dots, m_r $) denote the real eigenvalues, and $ \lambda_{c,i} $, $ \lambda_{c,i}^* $ ($ i = 1, 2, \dots, m_c $) represent the complex conjugate pairs, where $ \lambda_{c,i} = \sigma_i + j\omega_i $ with $ \sigma_i, \omega_i \in \mathbb{R} $ and $ \omega_i \neq 0 $. Here, $ m_r $ and $ m_c $ are the numbers of real and complex eigenvalues, respectively. These notations will be used to compactly express and derive the subsequent results in Section \ref{identification_algorithm}.

In general, the state transition matrix $\Xi$ can be arbitrary. However, for the sake of simplicity in derivation, we restrict our attention to the case where $\Xi$ has distinct eigenvalues.
\begin{assumption}
	\label{assum_Xi_eigenvalue}
	All eigenvalues of the state transition matrix $\Xi$ are distinct. 
\end{assumption}

For the matrix $\Xi$, there exists an invertible matrix $T$ and its Jordan canonical form $\Lambda$, such that $\Xi = T \Lambda T^{-1}$, defined as:
\begin{equation}
	\label{eq_jordon}
	\begin{aligned}
		& T = \operatorname{diag}\left\{ I_{m_r}, \underbrace{\begin{bmatrix} 1 & 1 \\ j & -j \end{bmatrix}, \dots, \begin{bmatrix} 1 & 1 \\ j & -j \end{bmatrix}}_{m_c \text{ blocks}} \right\} \\
		& \Lambda =\mathrm{diag}\left\{ \left. \lambda _{r,i} \right|_{i=1}^{m_r},\left. \left[ \begin{matrix}
			\lambda _{c,i}&		0\\
			0&		\lambda _{c,i}^{*}\\
		\end{matrix} \right] \right|_{i=1}^{m_c} \right\} 
	\end{aligned}
\end{equation}
Then, $\Pi T$ can be expressed as:
$$
\Pi T\triangleq  \begin{bmatrix}
	\pi _{r,1}&		\cdots&		\pi _{r,m_r}&		\pi _{c,1}&		\pi _{c,1}^{*}&		\cdots&		\pi _{c,m_c}&		\pi _{c,m_c}^{*}\\
\end{bmatrix} 
$$
where $\pi_{r,i}$ are real-valued vectors corresponding to the real eigenvalues, and $\pi_{c,i}$ are complex-valued vectors associated with the complex conjugate eigenvalue pairs.

\begin{remark}
	In fact, when the matrix $ \Xi $ has repeated eigenvalues, particularly when its algebraic multiplicity differs from its geometric multiplicity, not only the right tangential interpolations but also the derivative tangential interpolations of the system can be estimated from the input–output data. However, further efforts are still needed on how to utilize these derivative tangential interpolations for structure identification of the NDS.
\end{remark}

\subsection{Problem Formulation and Preliminaries}

\textbf{Identification Problem:} Given the known models of the NDS subsystems \eqref{eq:subsystem} and the input-generation system \eqref{eq_signal_generation}, and provided with a set of measurements of the external outputs from each subsystem, collected at its own set of asynchronous and potentially non-uniform and slow sampling instants, the objective is to estimate the vector of unknown interconnection parameters $\theta$ in \eqref{eq_Phi_theta}.

Notably, if there are some subsystem parameters to be estimated, they can be incorporated into the matrix $\Phi$ by introducing some virtual internal inputs and outputs, as detailed in \cite{Zhou2022a}. This implies that the results of this paper are also applicable to the parameter estimation of NDS subsystems.

To ensure the problem is well-posed, we adopt the following assumptions, which is necessary for open-loop experiments.
\begin{assumption}
	\label{assum_stability}
	The NDS $\Sigma_\theta$ is stable. An upper bound for its settling time, denoted as ${\bar{t}}_{s}$, is known. 
\end{assumption}

The following are some key results regarding the decomposition of system outputs and the definition of right tangential interpolation, which will be utilized in the derivation of the subsequent structure identification algorithm:

\begin{lemma}[Lemma 3 in \cite{Zhou2024a}]
	\label{lemma_output_decomposition}
	Under Assumption \ref{assum_1}, the external output $y(t)$ of the NDS $\Sigma_\theta$ can be explicitly decomposed into two parts, the transient response $y_t(t)$ and the steady-state response $ y_s(t) $, such that $ y(t) = y_t(t) + y_s(t) + n(t) $, if and only if there exist time-independent matrices $X, Z \in \mathbb{R}^{(m_x + m_z) \times m_\xi}$ satisfying the following two equations simultaneously:
	\begin{equation}
		\label{eq_Sylvester}
		\begin{matrix}
			\bar{E}X-Z=0,&		A_{\theta}X+\left[ \begin{array}{c}
				B_{xu}\\
				D_{zu}\\
			\end{array} \right] \Pi =Z\Xi\\
		\end{matrix}
	\end{equation}
	
	For any initial state vector $x(0)$ of the NDS $\Sigma_\theta$ and any initial state vector $\xi(0)$ of the input-generation  system $\Sigma_s$, the transient response $y_t(t)$ and steady-state response $y_s(t)$ are defined as follows:
	$$
	\begin{aligned}
		y_t(t)&=C_\theta\mathcal{L} ^{-1}\left\{ (s\bar{E}-A_\theta)^{-1} \right\} [\bar{E}x(0)-Z\xi (0)]\\
		y_s(t)&=\left(C_\theta X+D_{y u} \Pi\right) \xi(t)\\
	\end{aligned}
	$$
	
	Notably, $y_t(t)$ is induced by the initial states of the NDS $\Sigma_\theta$ and input-generation system $\Sigma_s$. When $\Sigma_\theta$ is stable, $y_t(t)$ decays exponentially to zero over time. The matrix $C_\theta X+D_{y u} \Pi$ is a constant matrix that only depends on the system matrices of $\Sigma_\theta$ and $\Sigma_s$. Under Assumption \ref{assum_Xi_eigenvalue}, if the eigenvalues of $\Xi$ and the generalized eigenvalues of the matrix pair $(\bar{E}, A_{\theta})$ are disjoint, then equation \eqref{eq_Sylvester} has a unique solution for $X$ and $Z$.
\end{lemma}

For the overall NDS system \( \Sigma_{\theta} \), the concept of right tangential interpolation characterizes the directional evaluation of its transfer function in the complex plane. The definition is given as follows.

\begin{definition}[Right Tangential Interpolation]
	\label{def_moment}
	Let \( H(s) \) denote the transfer function matrix of the NDS. For a point \( s \in \mathbb{C} \setminus \sigma(\bar{E}, A_{\theta}) \), where \( \sigma(\bar{E}, A_{\theta}) \) is the set of generalized eigenvalues of the system matrix pair $(\bar{E}, A_{\theta})$, and a prescribed right tangential direction \( \ell \in \mathbb{C}^{m_u} \), the right tangential interpolation of the system is defined as:
	\[
	\eta_k(s, \ell) = \frac{d^k H(s)}{ds^k}\,\ell \in \mathbb{C}^{m_y}, \quad k \in \mathbb{Z}_{\ge 0}.
	\]
	
	Although right tangential interpolations can, in general, be of arbitrary order, this paper only considers the zero‑order case, which is represented by $ \eta(s, \ell) = H(s)\,\ell $.

	In fact, the system’s tangential interpolations can be characterized through the Sylvester equation \eqref{eq_Sylvester}. The interpolation locations correspond to the eigenvalues of the input‑generation matrix \( \Xi \), while the tangential directions are determined by the output matrix \( \Pi \) together with the similarity transformation that brings \( \Xi \) to its Jordan canonical form.
\end{definition}

Lemma \ref{lemma_output_decomposition} indicates that the matrix \( (C_\theta X + D_{yu}\Pi) \), which governs the steady‑state response of the NDS, is directly related to the right tangential interpolations of the system, evaluated at the eigenvalues of the input generator matrix \( \Xi \). 

To formalize this connection, we define the following real matrices:
\begin{align*}
	R&=\mathrm{row}\left\{ \left. \eta \left( \lambda_{r,i},\pi_{r,i} \right) \right|_{i=1}^{m_r} \right\}\\
	C&=\mathrm{row}\left\{ \left. \begin{bmatrix}
		\eta \left( \lambda_{c,i},\pi_{c,i} \right)&		\eta \left( \lambda_{c,i}^{*},\pi_{c,i}^{*} \right)
	\end{bmatrix} \right|_{i=1}^{m_c} \right\}
\end{align*}
where $\eta(s,\ell)$ denotes the right tangential interpolation of the system transfer function at point $s$ along direction $\ell$. Consequently, we have the following corollary.

\begin{corollary}
	\label{corollary_zero_order_moment}
	Under Assumptions \ref{assum_1}–\ref{assum_Xi_eigenvalue}, the steady‑state response matrix can be expressed as:
	$$
	C_\theta X+D_{y u} \Pi=\begin{bmatrix}R&C\end{bmatrix}T^{-1}
	$$
	where $T$ is the transformation matrix defined in \eqref{eq_jordon}.
\end{corollary}

This relationship establishes a bridge between the system's time-domain behavior and its frequency-domain characteristics. Specifically, it shows that the steady-state response $y_s(t)$ encodes the system's frequency-domain characteristics at the interpolation points $\{\lambda_{r,i}\}_{i=1}^{m_r}$ and $\{\lambda_{c,i}, \lambda_{c,i}^*\}_{i=1}^{m_c}$ along the corresponding tangential directions $\{\pi_{r,i}\}_{i=1}^{m_r}$ and $\{\pi_{c,i}, \pi_{c,i}^*\}_{i=1}^{m_c}$. This enables the development of subsequent structure identification algorithm that can extract interconnection parameters from steady-state measurements, even when the data are collected under asynchronous, non-uniform, and slow-rate sampling conditions.

\section{Two-Stage Structure Identification Algorithm}
\label{identification_algorithm}

To address the challenge of identifying NDS from irregularly sampled data, we extend previous two-stage method \cite{Zhou2024a}. The first stage leverages the properties of the steady-state response to obtain a non-parametric estimate of the system's right tangential interpolations, effectively fusing all asynchronous measurements into a unified representation. The second stage, which constitutes the primary theoretical contribution of this work, introduces a left null‑space projection to eliminate the bilinear coupling between state‑related matrices and the structural parameters $\theta$. This decoupling transforms the bilinear problem into a sequence of linear estimation tasks, enabling robust and consistent reconstruction of $\theta$ from the interpolations.

\subsection*{Stage 1: Estimation of Right Tangential Interpolations from Irregular Data}
\label{sec:stage1}

This stage estimates the system’s right tangential interpolations from asynchronous time-domain output data. The key step is to rewrite the steady-state response as a linear regression model whose unknowns are the real and imaginary parts of the interpolations. This formulation lets us combine all samples from different subsystems and sampling times into one regression to obtain a non-parametric estimate of the interpolations.

The starting point is the explicit formula for the steady-state response, which we present as the following theorem, adapted from Corollary 1 and the surrounding derivations in our preliminary work \cite{Zhou2024a}.

For brevity, we use $\eta_{\star,i}$ to denote $\eta(\lambda_{\star,i}, \pi_{\star,i})$, where $\star \in \{r,c\}$ and $i = 1, \dots, m_\star$. Here, $\lambda_{\star,i}$ denotes the $i$-th real ($\star = r$) or complex ($\star = c$) eigenvalue of $\Xi$, and $\pi_{\star,i}$ is the corresponding direction vector.

\begin{theorem}[Steady-State Response Decomposition]
	\label{thm:steady_state_formula}
	Under Assumptions \ref{assum_1} and \ref{assum_Xi_eigenvalue}, the steady-state response $y_s(t)$ of the NDS $\Sigma_\theta$ can be expressed as:
	\begin{equation}
		\label{eq:steady_state_response_formula}
		y_s(t)=\sum_{i=1}^{m_r}{\alpha _ie^{\lambda _{r,i}t}\eta _{r,i}}+\sum_{i=1}^{m_c}{\mathrm{Re}\left\{ \beta _{i}^{*}e^{\lambda _{c,i}t}\eta _{c,i} \right\}}
	\end{equation}
	\begin{proof}
		The proof is provided in the Appendix.
	\end{proof}
\end{theorem}

This theorem reveals that $y_s(t)$ is a linear combination of the system's right tangential interpolations. The coefficients in this combination, the real scalars $\alpha_i$ and complex scalars $\beta_i = \mu_i + j\nu_i$, are directly determined by the choice of the initial state $\xi(0)$ for the input-generation system $\Sigma_s$. We can structure the initial state vector $\xi(0)$ to directly set the input signal components:
\begin{equation}
	\label{eq_xi_0}
	\xi(0) = \operatorname{col}\left\{ \left. \alpha_i \right|_{i=1}^{m_r}, \left. \begin{bmatrix} \mu_i \\ \nu_i \end{bmatrix} \right|_{i=1}^{m_c} \right\}
\end{equation}
where $\alpha_i$ determines the amplitude of the real exponential term $e^{\lambda_{r,i}t}$, while $\mu_i$ and $\nu_i$ jointly determine the amplitude and phase of the complex exponential term $e^{\lambda_{c,i}t}$.

To formulate the linear regression problem, we introduce a real‑valued parameter vector $\bar{\eta}$ that collects all unknown interpolations within a single vector.
\begin{definition}[Interpolation Vector]
	The vector of right tangential interpolations to be estimated is defined as:
	\begin{equation}
		\label{eq_eta_bar_detailed}
		\bar{\eta}=\mathrm{col}\left\{ \left. \eta _{r,i} \right|_{i=1}^{m_r},\left. \left[ \begin{array}{c}
			\mathrm{Re}\{\eta _{c,i}\}\\
			\mathrm{Im}\{\eta _{c,i}\}\\
		\end{array} \right] \right|_{i=1}^{m_c} \right\}
	\end{equation}
	where $\bar{\eta} \in \mathbb{R}^{(m_r+2m_c)m_y}$.
\end{definition}

Next, we construct a corresponding real-valued regressor vector. Let $\phi_i = \arctan(\nu_i/\mu_i)$. For each complex eigenvalue pair $\lambda_{c,i} = \sigma_i + j\omega_i$, the complex term in \eqref{eq:steady_state_response_formula} can be expanded as:
$$
\begin{aligned}
	\mathrm{Re}\left\{ \beta _{i}^{*}e^{\lambda _{c,i}t}\eta _{c,i} \right\} =&\sqrt{\mu _{i}^{2}+\nu _{i}^{2}}e^{\sigma _it}\cos\mathrm{(}\omega _it+\phi _i)\mathrm{Re}\{\eta _{c,i}\}\\
	&-\sqrt{\mu _{i}^{2}+\nu _{i}^{2}}e^{\sigma _it}\sin\mathrm{(}\omega _it+\phi _i)\mathrm{Im}\{\eta _{c,i}\}\\
\end{aligned}
$$

This reveals the real-valued, time-varying coefficients associated with the real and imaginary components of each complex interpolation.

\begin{definition}[Regressor Vector]
	The time-varying regressor vector is defined by collecting all scalar coefficients from the expanded steady-state response:
	$$
	\begin{aligned}
		\label{eq_psi_detailed}
		\psi(t)=\begin{bmatrix}
			\mathrm{col}\left\{\left.\alpha _i e^{\lambda _{r,i}t}\right|_{i = 1}^{m_r}\right\} \\
			\mathrm{col}\left\{\left.\begin{bmatrix}
				\sqrt{\mu _{i}^{2}+\nu _{i}^{2}}\cos\mathrm{(}\omega _it+\phi _i)e^{\sigma _it} \\
				-\sqrt{\mu _{i}^{2}+\nu _{i}^{2}}\sin\mathrm{(}\omega _it+\phi _i)e^{\sigma _it}
			\end{bmatrix}\right|_{i = 1}^{m_c}\right\}
		\end{bmatrix}\
	\end{aligned}
	$$
	where $\psi(t) \in \mathbb{R}^{ (m_r+2m_c)}$.
\end{definition}

The crucial step for handling asynchronicity is to construct a specific regressor matrix for each individual data sample. Let ${t}_{j}^{\left\lbrack k \right\rbrack}$ denote the $j$-th sampling instant for the $k$-th subsystem. The measured output $y_k(t_j^{[k]})$ is a subvector of the total system output $y(t_j^{[k]})$. We can extract it using a selection matrix  defined as: \(S_k = \begin{bmatrix} 0 & \dots & 0 & I_{m_{y,k}} & 0 & \dots & 0 \end{bmatrix} \in \mathbb{R}^{m_{y,k} \times m_y}\), where the identity matrix $I_{m_{y,k}}$ is positioned to select the rows corresponding to the $k$-th subsystem's output. Based on these notations, the following corollary holds, which provides a linear regression for any single asynchronous sample.

\begin{corollary}
	\label{cor:single_sample_model}
	Under Assumptions \ref{assum_1}-\ref{assum_Xi_eigenvalue}, the external output of the $k$-th subsystem at the sampling instant ${t}_{j}^{\left\lbrack k \right\rbrack}$ is given by
	\begin{equation}
		\label{eq:single_sample_model}
		{y}_{k}\left( {t}_{j}^{\left\lbrack  k\right\rbrack  }\right)  = \Gamma \left( {t}_{j}^{\left\lbrack  k\right\rbrack  }\right) \bar{\eta } + {y}_{t,k}\left( {t}_{j}^{\left\lbrack  k\right\rbrack  }\right)  + {n}_{k}\left( {t}_{j}^{\left\lbrack  k\right\rbrack  }\right)
	\end{equation}
	where ${y}_{t,k}$ and ${n}_{k}$ are the transient and noise components for the $k$-th subsystem, and the regressor matrix $\Gamma \left( {t}_{j}^{\left\lbrack  k\right\rbrack  }\right) \in \mathbb{R}^{m_{y,k} \times (m_r+2m_c)m_y}$ is defined as:
	\begin{equation}
		\Gamma \left( {t}_{j}^{\left\lbrack  k\right\rbrack  }\right)  = S_k (\psi(t_j^{[k]})^T \otimes I_{m_y})
	\end{equation}
	\begin{proof}
		Owing to space limits and the straightforward derivation, the proof is omitted.
	\end{proof}
\end{corollary}

The matrix $\Gamma(t_j^{[k]}) \in \mathbb{R}^{m_{y,k} \times (m_r+2m_c)m_y}$ is the regressor for this specific measurement. It provides a linear constraint on the global interpolation vector $\bar{\eta}$, thereby fusing this local, asynchronous piece of information into a single coherent estimation problem.

To perform the estimation, we collect all sampling instants across all subsystems and arrange them in nondecreasing order as $T_1, \ldots, T_{m_t}$, without assuming uniform spacing. Let $y_m^{T_i}$ be the measured external output at $T_i$, $y_t^{T_i}$ its transient part, and $n^{T_i}$ the measurement noise. To facilitate the derivations, the following vectors and matrices are defined:
\begin{equation}
	\label{eq_step1}
	\begin{aligned}
		y_m\!\left( T_{1:m_t} \right) &=\mathrm{col}\!\left\{ \left. y_{m}^{T_i} \right|_{i=1}^{m_t} \right\}\\
		n\left( T_{1:m_t} \right) &=\mathrm{col}\!\left\{ \left. n^{T_i} \right|_{i=1}^{m_t} \right\}\\
		\Gamma \!\left( T_{1:m_t} \right) &=\mathrm{row}\!\left\{ \left. \Gamma (T_i) \right|_{i=1}^{m_t} \right\}\\
	\end{aligned}
\end{equation}

Under Assumptions \ref{assum_1}-\ref{assum_stability}, the transient $y_t(t)$ decays exponentially. Let $m_0$ be the first index with $T_{m_0}\ge \bar{t}_s$, where $\bar{t}_s$ is the  upper bound on the settling time. Using only steady-state samples $T_{m_0},\ldots,T_{m_t}$, the least-squares estimate is
\begin{equation}
	\label{eq_eta_hat_detailed}
	\widehat{\bar{\eta}}=\Gamma \!\left( T_{m_0:m_t} \right) ^{\dagger}y_m\!\left( T_{m_0:m_t} \right) 
\end{equation}
under the condition that $\Gamma\!\left( T_{m_0:m_t} \right)$ has full column rank.

For online updates when new steady-state samples arrive, given initial values $\hat{\bar{\eta}}(0)$ and $P(0)$, the Recursive least-squares (RLS) recursion is
$$
\begin{aligned}
	& \hat{\bar{\eta}}(k)=\hat{\bar{\eta}}(k-1)+K(k)\left[ y_{m}(T_{m_0+k})-\varGamma (T_{m_0+k})\hat{\bar{\eta}}(k-1) \right] \\
	& \begin{aligned}
		K(k)= & P(k-1)\varGamma ^T(T_{m_0+k})                                                   \\
		& \cdot \left[ \varGamma (T_{m_0+k})P(k-1)\varGamma ^T(T_{m_0+k})+I \right] ^{-1} \\
	\end{aligned}             \\
	& P(k)=\left[ I-K(k)\varGamma (T_{m_0+k}) \right] P(k-1)                                                                    \\
\end{aligned}
$$
for $k = 1,2,\ldots,m_t-m_0+1$, where $I$ is an identity matrix of compatible dimension. Here, $\hat{\bar{\eta}}(k)$ is the estimate after processing the $k$-th steady-state sample, and $K(k)$ and $P(k)$ are the gain and covariance matrices. This is a standard result; see \cite{Ljung1999} for properties such as convergence.

\begin{remark}
	\label{identi_stage_1}
	The matrix $\Gamma\!\left( T_{m_0:m_t} \right)$ has full column rank if the regression is persistently excited. In particular, this holds if all components of $\xi(0)$ are nonzero and the number of steady-state samples exceeds $m_\xi$, the number of eigenvalues of $\Xi$ (which equals the number of right tangential interpolations to be estimated). This requirement is relatively mild and can be satisfied in most practical scenarios.
\end{remark}

\subsection*{Stage 2: Parameter Reconstruction via Algebraic Decoupling}
\label{sec:stage2}

With the estimated interpolation vector $\widehat{\bar{\eta}}$ from Stage 1, we proceed to recover the parameter vector $\theta$. The core challenge is that the underlying unknowns, the intermediate matrix $X$ from Lemma \ref{lemma_output_decomposition} and the parameter vector $\theta$, are coupled in a bilinear manner. Our approach systematically decouples this bilinearity, transforming the nonlinear problem into a sequence of two linear estimation problems.

First, we reconstruct the estimated steady-state response matrix, $\widehat{Y}_{ss}$, from the interpolation vector $\widehat{\bar{\eta}}$. From Corollary \ref{corollary_zero_order_moment}, we know $Y_{ss} = C_\theta X + D_{yu}\Pi = \begin{bmatrix}R&C\end{bmatrix}T^{-1}$. The matrix $\begin{bmatrix}R&C\end{bmatrix}$ is composed of the right tangential interpolations. We can construct its estimate, $\begin{bmatrix}\widehat{R}&\widehat{C}\end{bmatrix}$, by reshaping the estimated interpolation vector $\widehat{\bar{\eta}}$. Thus, the estimate for the steady-state response matrix is $\widehat{Y}_{ss} = \begin{bmatrix}\widehat{R}&\widehat{C}\end{bmatrix}T^{-1}$.

The relationship between the unknowns ($X, \theta$) and the knowns (system matrices and $\widehat{Y}_{ss}$) is defined by three fundamental matrix equations derived from Lemma \ref{lemma_output_decomposition}. Let the matrix $X \in \mathbb{R}^{(m_x + m_z) \times m_\xi}$ be partitioned as $X = \begin{bmatrix} X_{top}^T & X_{btm}^T \end{bmatrix}^T$, where $X_{top} \in \mathbb{R}^{m_x \times m_\xi}$ corresponds to the states $x(t)$, and $X_{btm} \in \mathbb{R}^{m_z \times m_\xi}$ corresponds to the internal outputs $z(t)$. The governing equations are:
\begin{enumerate}
	\item \textbf{Dynamics Equation:}
	\begin{equation} \label{eq:dyn_ss}
		A_{xx}X_{top} + B_{xv}\Phi(\theta)X_{btm} + B_{xu}\Pi = EX_{top}\Xi
	\end{equation}
	\item \textbf{Internal Constraint Equation:}
	\begin{equation} \label{eq:constraint_ss}
		C_{zx}X_{top} + D_{zv}\Phi(\theta)X_{btm} + D_{zu}\Pi = X_{btm}
	\end{equation}
	\item \textbf{External Output Equation:}
	\begin{equation} \label{eq:output_ss}
		C_{yx}X_{top} + D_{yv}\Phi(\theta)X_{btm} + D_{yu}\Pi = \widehat{Y}_{ss}
	\end{equation}
\end{enumerate}

These equations are bilinear in the unknowns $X$ and $\theta$, i.e., they include the product $\Phi(\theta)X_{btm}$. Our approach decouples this structure by first solving a linear problem for X and then estimating $\theta$.

\subsubsection*{Step 2a: Linear Estimation of the Intermediate Matrix X}

To isolate $X$, we must eliminate all terms containing the unknown parameter vector $\theta$. We achieve this via left null-space projection. We use the affine expansion $\Phi(\theta) = \Phi_0 + \sum_{i=1}^{m_\theta} \theta_i \Phi_i$, where $\Phi_0$ is a known constant matrix.
\begin{definition}[Parameter Coefficient Matrices]
	Based on the affine structure, we define:
	\begin{equation*}
		\begin{aligned}
			M & := [B_{xv}\Phi_1, B_{xv}\Phi_2, \dots, B_{xv}\Phi_{m_\theta}] \in \mathbb{R}^{m_x \times (m_\theta m_z)} \\
			N & := [D_{zv}\Phi_1, D_{zv}\Phi_2, \dots, D_{zv}\Phi_{m_\theta}] \in \mathbb{R}^{m_z \times (m_\theta m_z)} \\
			Q & := [D_{yv}\Phi_1, D_{yv}\Phi_2, \dots, D_{yv}\Phi_{m_\theta}] \in \mathbb{R}^{m_y \times (m_\theta m_z)}
		\end{aligned}	
	\end{equation*}
	
	For each of these matrices, we compute its Singular Value Decomposition. For instance, for $M$, we have $M = U_M \Sigma_M V_M^\top$. We then partition the matrix of left singular vectors $U_M$ as $U_M = [U_{M,1} \ | \ U_{M,2}]$, where the columns of $U_{M,1}$ form an orthonormal basis for the range space of $M$, and the columns of $U_{M,2}$ form an orthonormal basis for its left null-space. By definition, $U_{M,2}^\top M = 0$.
\end{definition}

\begin{remark}
	For networked systems, the number of internal inputs and outputs, $m_v$ and $m_z$, is typically much larger than the number of external inputs and outputs. Furthermore, the interconnection topology is often sparse, meaning the matrices $\Phi_i$ that define the structural basis are themselves sparse and of low rank. Consequently, the coefficient matrices $M, N, Q$, which are constructed from these $\Phi_i$ matrices, are generally rank-deficient. This guarantees the existence of a non-trivial left null-space, making the proposed projection method widely applicable to practical large-scale NDS.
\end{remark}

By left-multiplying equations \eqref{eq:dyn_ss}-\eqref{eq:output_ss} with the corresponding null-space matrices ($U_{M,2}^\top$, $U_{N,2}^\top$, $U_{Q,2}^\top$), all terms involving the unknown $\theta_i$ ($i=1,\dots,m_\theta$) are eliminated, but the constant $\Phi_0$ term remains. This procedure yields a set of linear equations solely in terms of $X_{top}$ and $X_{btm}$:
\begin{align*}
	U_{M,2}^\top (A_{xx}X_{top} + B_{xv} \Phi_0 X_{btm} + B_{xu}\Pi)  & = U_{M,2}^\top EX_{top}\Xi      \\
	U_{N,2}^\top (C_{zx}X_{top} + D_{zv} \Phi_0 X_{btm} + D_{zu}\Pi)  & = U_{N,2}^\top X_{btm}          \\
	U_{Q,2}^\top (C_{yx}X_{top} + D_{yv} \Phi_0 X_{btm} + D_{yu}\Pi ) & = U_{Q,2}^\top \widehat{Y}_{ss}
\end{align*}

These three matrix equations are now linear in the unknowns $X_{top}$ and $X_{btm}$. To solve for them, we vectorize the equations using the property $\operatorname{vec}(ABC) = (C^T \otimes A)\operatorname{vec}(B)$. This allows us to assemble a single large linear system.
Let $x = \begin{bmatrix} \operatorname{vec}(X_{top})^T & \operatorname{vec}(X_{btm})^T \end{bmatrix}^T$. The system takes the form:
\begin{equation*}
	\varGamma x = \gamma
\end{equation*}
where the system matrix $\varGamma$ and vector $\gamma$ are defined as:
\begin{equation}
	\label{eq:matrix_A_blocks}
	\varGamma =
	\begin{bmatrix}
		\varGamma_{11} & \varGamma_{12} \\
		\varGamma_{21} & \varGamma_{22} \\
		\varGamma_{31} & \varGamma_{32}
	\end{bmatrix}
\end{equation}
with blocks
\begin{align*}
	\varGamma_{11} & = I_{m_\xi} \otimes (U_{M,2}^\top A_{xx}) - \Xi^T \otimes (U_{M,2}^\top E) \\
	\varGamma_{12} & = I_{m_\xi} \otimes (U_{M,2}^\top B_{xv} \Phi_0)                                     \\
	\varGamma_{21} & = I_{m_\xi} \otimes (U_{N,2}^\top C_{zx})                   \\                         
	\varGamma_{22} & = I_{m_\xi} \otimes (U_{N,2}^\top (D_{zv} \Phi_0 - I_{m_z}))                         \\
	\varGamma_{31} & = I_{m_\xi} \otimes (U_{Q,2}^\top C_{yx})                                            \\
	\varGamma_{32} & = I_{m_\xi} \otimes (U_{Q,2}^\top D_{yv} \Phi_0)
\end{align*}
and
\begin{equation*}
	\gamma =
	\begin{bmatrix}
		-\operatorname{vec}(U_{M,2}^\top B_{xu}\Pi) \\
		-\operatorname{vec}(U_{N,2}^\top D_{zu}\Pi) \\
		\operatorname{vec}(U_{Q,2}^\top (\widehat{Y}_{ss} - D_{yu}\Pi))
	\end{bmatrix}
\end{equation*}

If $\varGamma$ has full column rank, we can obtain a unique least-squares estimate for the vectorized intermediate matrix $X$:
\begin{equation} \label{eq:X_hat_final}
	\widehat{x} = \begin{bmatrix} \operatorname{vec}(\widehat{X}_{top}) \\ \operatorname{vec}(\widehat{X}_{btm}) \end{bmatrix} = \varGamma^\dagger \gamma
\end{equation}
The matrices $\widehat{X}_{top}$ and $\widehat{X}_{btm}$ are then recovered by reshaping.

\subsubsection*{Step 2b: Linear Estimation of the Structure Parameter $\theta$}
With the estimate $\widehat{X}$ in hand, we return to the original equations \eqref{eq:dyn_ss}-\eqref{eq:output_ss}. Since $X$ is now treated as known (by substituting $\widehat{X}$), these equations become linear in the unknown parameter vector $\theta$. Using the affine expansion $\Phi(\theta) = \Phi_0 + \sum_{i=1}^{m_\theta} \theta_i \Phi_i$, we rearrange each equation to isolate the terms containing $\theta$:
\begin{equation}
	\label{eq:theta_lin_1}
	\begin{aligned}
	B_{xv}\left(\sum_{i=1}^{m_\theta}\theta_i\Phi_i\right)\widehat{X}_{btm} = & E\widehat{X}_{top}\Xi - A_{xx}\widehat{X}_{top} \\
	& - B_{xv}\Phi_0\widehat{X}_{btm} - B_{xu}\Pi
	\end{aligned}
\end{equation}

\begin{align}
	\label{eq:theta_lin_2}
	\begin{aligned}
		D_{zv}\left(\sum_{i=1}^{m_\theta}\theta_i\Phi_i\right)\widehat{X}_{btm} = & \widehat{X}_{btm} - C_{zx}\widehat{X}_{top} \\
		& - D_{zv}\Phi_0\widehat{X}_{btm} - D_{zu}\Pi
	\end{aligned}     \\
	\label{eq:theta_lin_3}
	\begin{aligned}
		D_{yv}\left(\sum_{i=1}^{m_\theta}\theta_i\Phi_i\right)\widehat{X}_{btm} = & \widehat{Y}_{ss} - C_{yx}\widehat{X}_{top}  \\
		& - D_{yv}\Phi_0\widehat{X}_{btm} - D_{yu}\Pi
	\end{aligned}
\end{align}

Each of these equations is now linear in the parameter vector $\theta=[\theta_1, \dots, \theta_{m_\theta}]^T$. To construct a single linear system, we vectorize both sides.

Regressor Matrix $\Psi (\widehat{X}_{btm})$ and Observation Vector $\kappa(\widehat{X}_{btm})$. Let $\Psi_1, \Psi_2, \Psi_3$ be matrices whose $i$-th column is given by:
\begin{align*}
	\operatorname{col}_i(\Psi_1) & = \operatorname{vec}(B_{xv}\Phi_i\widehat{X}_{btm}) \\
	\operatorname{col}_i(\Psi_2) & = \operatorname{vec}(D_{zv}\Phi_i\widehat{X}_{btm}) \\
	\operatorname{col}_i(\Psi_3) & = \operatorname{vec}(D_{yv}\Phi_i\widehat{X}_{btm})
\end{align*}

Let $\kappa_1, \kappa_2, \kappa_3$ be vectors corresponding to the vectorized right-hand sides of \eqref{eq:theta_lin_1}-\eqref{eq:theta_lin_3}:
\begin{align*}
	\kappa_1 & = \operatorname{vec}(E\widehat{X}_{top}\Xi - A_{xx}\widehat{X}_{top} - B_{xv}\Phi_0\widehat{X}_{btm} - B_{xu}\Pi) \\
	\kappa_2 & = \operatorname{vec}(\widehat{X}_{btm} - C_{zx}\widehat{X}_{top} - D_{zv}\Phi_0\widehat{X}_{btm} - D_{zu}\Pi)     \\
	\kappa_3 & = \operatorname{vec}(\widehat{Y}_{ss} - C_{yx}\widehat{X}_{top} - D_{yv}\Phi_0\widehat{X}_{btm} - D_{yu}\Pi)
\end{align*}

The final regressor matrix $\Psi$ and observation vector $\kappa$ are constructed by stacking these components:
\begin{equation}
	\label{eq:psi_kappa_def}
	\begin{matrix}
		\Psi (\widehat{X}_{btm})=\left[ \begin{array}{c}
			\Psi _1\\
			\Psi _2\\
			\Psi _3\\
		\end{array} \right] ,&		\kappa (\widehat{X},\widehat{Y}_{ss})=\left[ \begin{array}{c}
			\kappa _1\\
			\kappa _2\\
			\kappa _3\\
		\end{array} \right]\\
	\end{matrix}
\end{equation}

The three vectorized equations can now be written compactly as a single linear system:
\begin{equation*}
	\Psi (\widehat{X}_{btm}) \theta = \kappa (\widehat{X},\widehat{Y}_{ss})
\end{equation*}

The final least-squares estimate for the interconnection parameters is then given by:
\begin{equation} \label{eq:theta_hat_final}
	\widehat{\theta }=\Psi (\widehat{X}_{btm})^{\dagger}\kappa (\widehat{X},\widehat{Y}_{ss})
\end{equation}

\begin{remark}
	It is important to note that the proposed two-stage algorithm involves a propagation of estimation errors. The estimate of the intermediate matrix, $\widehat{X}$, is computed in Step 2a using the estimated steady-state response matrix, $\widehat{Y}_{ss}$, which is derived from noisy data. Subsequently, the final parameter estimate, $\widehat{\theta}$, is calculated in Step 2b using both $\widehat{X}$ and $\widehat{Y}_{ss}$. This sequential dependency means that the accuracy of $\widehat{\theta}$ is sensitive to the estimation errors from the preceding steps. This sensitivity is a characteristic trade-off for transforming a non-convex bilinear problem into a solvable sequence of linear problems, and its effects will be illustrated in the numerical simulations in Section \ref{sec:analysis}.
\end{remark}

\begin{remark}
	This paper presents two major extensions over the framework in \cite{Zhou2024a}: extending its application to asynchronously sampled networked systems rather than a single lumped system, and relaxing the identifiability conditions. The approach in \cite{Zhou2024a} required certain transfer function matrices to have full normal rank, the method presented here bypasses this prerequisite. However, this condition is not entirely eliminated but is replaced by a different requirement: for the intermediate matrix $X$ to be uniquely solvable in Step 2a, the matrix $\varGamma$ in \eqref{eq:matrix_A_blocks} must have full column rank. This new condition serves as the identifiability and persistent excitation requirement for the algorithm. It implies that the number and placement of external inputs and outputs must be rich enough to ensure that the system's internal dynamics, encapsulated in $X$, can be uniquely recovered from the external measurements. These changes allow the method to apply to a much broader class of networked dynamical systems, as illustrated by the simulation example in Section \ref{simulation}.
\end{remark}

\section{Theoretical Analysis}
\label{sec:analysis}

In this section, we provide a theoretical analysis of the proposed two-stage identification algorithm. We first discuss the conditions for persistent excitation and identifiability, which guarantee the existence of unique solutions in both stages of the estimation. Subsequently, we investigate the asymptotic properties of the estimators, such as their unbiasedness and consistency, as the number of data samples grows.

\subsection{Persistent Excitation and Identifiability Conditions}

Persistent excitation ensures that the input signals are sufficiently rich to uniquely determine the unknown parameters from the output data. In our framework, this translates to rank conditions on the regressor matrices constructed in each stage.

\subsubsection{Identifiability of Right Tangential Interpolations (Stage 1)}
The least-squares estimate $\widehat{\bar{\eta}}$ in \eqref{eq_eta_hat_detailed} is unique if and only if the matrix $\Gamma\!\left( T_{m_0:m_t} \right)$ has full column rank. This serves as the persistent excitation condition for the first stage. As discussed in Remark \ref{identi_stage_1}, this condition is generally satisfied if two requirements are met:
\begin{enumerate}
	\item For each distinct eigenvalue of the matrix $\Xi$, the corresponding component of the initial state $\xi(0)$ is non-zero, i.e. $\alpha_i \neq 0$ and $\mu_i, \nu_i$ are not simultaneously zero. This condition ensures that all modes associated with the eigenvalues of $\Xi$ are excited.
	\item The number of steady-state data samples used for estimation, $m_p = m_t - m_0 + 1$, is sufficiently large, and the sampling instants $T_{m_0}, \dots, T_{m_t}$ are sufficiently diverse to ensure linear independence among the columns of the regressor matrix $\Gamma \!\left( T_{m_0:m_t} \right)$.
\end{enumerate}

In practice, using an input signal composed of multiple sinusoids and ensuring a sufficient number of measurements typically satisfies this condition, which is a relatively mild condition that is typically easy to satisfy in practice.

\subsubsection{Identifiability of System Parameters (Stage 2)}
The identifiability in Stage 2 depends on the unique solvability of the two sequential linear systems.

\textbf{Identifiability of the Intermediate Matrix X:} The estimate $\widehat{x}$ in \eqref{eq:X_hat_final} is uniquely determined if and only if the matrix $\varGamma$ defined in \eqref{eq:matrix_A_blocks} has full column rank. This condition is the key identifiability requirement for the intermediate stage of our algorithm. The full rank of $\varGamma$ depends on the system's structure ($A_{xx}, E, C_{zx}$, etc.), the choice of input signal (through $\Xi$), and the null-space projection matrices ($U_{M,2}, U_{N,2}, U_{Q,2}$). Intuitively, it means that the available external measurements, after being projected to remove the influence of $\theta$, must contain enough information to uniquely determine the internal steady-state behavior represented by $X$.

\textbf{Identifiability of the Parameter Vector $\theta$:} Given a unique estimate $\widehat{X}$, the final parameter estimate $\widehat{\theta}$ in \eqref{eq:theta_hat_final} is unique if and only if the regressor matrix $\Psi (\widehat{X}_{btm})$ defined in \eqref{eq:psi_kappa_def} has full column rank. The matrix $\Psi (\widehat{X}_{btm})$ is a function of the known system matrices and the estimated intermediate matrix $\widehat{X}$. The condition that $\Psi (\widehat{X}_{btm})$ has full column rank is the final persistent excitation condition. It ensures that, once the internal steady-state behavior is known, the way the parameters $\theta_i$ influence the system equations is independent, allowing each $\theta_i$ to be uniquely distinguished. Importantly, this identifiability condition depends solely on the inherent structural properties of the NDS $\Sigma_\theta$ itself, rather than on the identification algorithm or external data characteristics.

\subsection{Asymptotic Properties of the Estimators}

We now analyze the statistical properties of the estimators as the number of steady-state data samples, $m_p$, tends to infinity. We make the standard assumption on the measurement noise.

\begin{assumption}[Noise Properties]
	\label{assum:noise}
	The noise sequence $n_k(t_j^{[k]})$ is a zero-mean stochastic process, uncorrelated over time interval and across subsystems. There exists a finite constant $\sigma_n^2$ such that the covariance of the stacked noise vector $n$ in \eqref{eq_step1} satisfies $E[n n^T] \leq \sigma_n^2 I$.
\end{assumption}

\begin{theorem}[Consistency of the Interpolation Estimator]
	\label{thm:consistency_stage1}
	Under Assumption \ref{assum:noise}, if the input is persistently exciting, the interpolation estimate $\widehat{\bar{\eta}}$ in \eqref{eq_eta_hat_detailed} is mean-square consistent, which implies it is asymptotically unbiased and consistent in probability. That is,
	\begin{enumerate}
		\item $E[\widehat{\bar{\eta}}] \to \bar{\eta}$ as $m_p \to \infty$.
		\item $\lim_{m_p \to \infty} E[||\widehat{\bar{\eta}} - \bar{\eta}||^2] = 0$.
	\end{enumerate}
\end{theorem}
\begin{proof}
	The proof follows standard least-squares analysis and is omitted.
\end{proof}

The consistency of the interpolation estimate $\widehat{\bar{\eta}}$ is the foundation for the consistency of the final parameter estimate. Since $\widehat{Y}_{ss}$ is a linear transformation of $\widehat{\bar{\eta}}$, and the estimation of $\widehat{X}$ is a linear transformation of $\widehat{Y}_{ss}$, the consistency of $\widehat{\bar{\eta}}$ implies the consistency of $\widehat{Y}_{ss}$ and $\widehat{X}$.

\begin{theorem}[Consistency of Parameter Estimator]
	\label{thm:consistency}
	When the identifiability conditions are satisfied, that is, the matrix $\Gamma$ defined in \eqref{eq:matrix_A_blocks} has full column rank, and the regressor matrix $\Psi(X)$ defined in \eqref{eq:psi_kappa_def} evaluated at the true parameter has full column rank, the algorithm is consistent, i.e., $\widehat{\theta} \to \theta$ in probability.
\end{theorem}
\begin{proof}
	The proof is provided in the Appendix.
\end{proof}

\begin{theorem}[Asymptotic Unbiasedness]
	\label{thm:ms_consistency}
	In addition to the assumptions of Theorem \ref{thm:consistency}, suppose that there exists a constant $C_{\Psi}<\infty$ such that for all sufficiently large $m_p$,
	$$
	\sup_{m_p} E\big[\,\sigma_{\min}(\Psi(\widehat{X}))^{-4}\,\big] \le C_{\Psi}.
	$$
	Then the identification algorithm of parameter $\theta$ is mean-square consistent and asymptotically unbiased.
\end{theorem}
\begin{proof}
	The proof is provided in the Appendix.
\end{proof}

Theorems \ref{thm:consistency_stage1}, \ref{thm:consistency}, and \ref{thm:ms_consistency} provide strong theoretical guarantees for the proposed algorithm. They ensure that with a sufficiently large number of steady-state measurements, the estimated interpolations and, consequently, the estimated structural parameters will converge to their true values, despite the presence of measurement noise and the complexity of the data irregularities.

\section{Numerical Simulations}
\label{simulation}

\begin{figure}[htbp]
	\centering
	\includegraphics[width=0.5\textwidth]{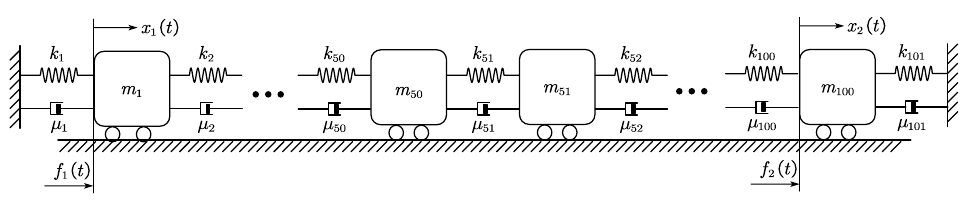}
	\caption{A spring-mass-damper system with 100 carts.}
	\label{fig:system_100}
\end{figure}

To verify the effectiveness of the proposed two-stage algorithm, this section presents numerical simulations on a spring-mass-damper system consisting of 100 carts connected in series by springs and dampers, as shown in Fig.~\ref{fig:system_100}. This system is specifically designed to be unsolvable by previous methods \cite{Zhou2024a}. Furthermore, we compare the proposed method with the traditional nonlinear least-squares (NLS) approach. The results demonstrate that our algorithm not only accurately recovers the interconnection structure of the system but also successfully overcomes the limitations of previous methods and the local minimum challenges faced by the NLS approach.

\subsection{Simulation Setup}

\textbf{Parameter Selection:} The system parameters are randomly generated: masses $m_i$ are drawn from a uniform distribution in $[1, 1.5]$, spring constants $k_i$ from $[0.5, 2.0]$, and damping coefficients $\mu_i$ from $[0.1, 0.5]$. To test the algorithm's capability, most connection parameters are assumed to be known, while only a pair from the middle of the system (e.g., $k_{51}$ and $\mu_{51}$) are treated as unknown parameters to be estimated.

\textbf{Input Signal:} The input signal is generated by a first-order oscillatory signal generator $\Sigma_s$ with a frequency of $\lambda = 0.32$ rad/s, applying external forces $f_1(t)$ and $f_2(t)$ to the first and last carts, respectively. The dynamics of the signal generator are characterized by:
$$
\Xi =\left[ \begin{matrix}
	0&		0.32\\
	-0.32&		0\\
\end{matrix} \right] ,\Pi =\left[ \begin{matrix}
	1.5&		2\\
	2&		1\\
\end{matrix} \right] , \xi \left( 0 \right) =\left[ \begin{array}{c}
	1\\
	1\\
\end{array} \right] 
$$

\textbf{Configuration:} In this system, a two-dimensional external excitation $u(t)$ is applied to the first and last carts, and only their positions are measured as the two-dimensional external output $y(t)$. The internal input $v(t)$ corresponds to the interaction forces between carts generated by springs and dampers, while the internal output $z(t)$ consists of the positions and velocities of all carts, both having a dimension of 200.

This results in neither of $G_{yv}$ and $G_{zu}$ can satisfy the full-rank condition. Therefore, previous method \cite{Zhou2024a} is inapplicable to such a large-scale system with a sparse configuration of sensors and actuators. In contrast, the new algorithm proposed in this paper bypasses this restriction through an algebraic decoupling technique, demonstrating its broader applicability.

\textbf{Sampling and Noise:} To simulate a challenging practical measurement environment, the outputs are sampled asynchronously, with sampling intervals chosen randomly between $0.1\,\mathrm{s}$ and $5\,\mathrm{s}$. This setup results in asynchronous, non-uniform, and slow-rate (partially sub-Nyquist) sampling conditions. The upper bound for the steady-state settling time is $\bar{t}_s = 14.25 \, \mathrm{s}$. The measured outputs are corrupted by additive Gaussian white noise with a variance of $\sigma_n=0.3$. The maximum amplitudes of the two outputs are 1.1182 and 1.9068, respectively.

\subsection{Simulation Results and Analysis}

\begin{figure}[htbp]
	\centering
	\includegraphics[width=0.35\textwidth]{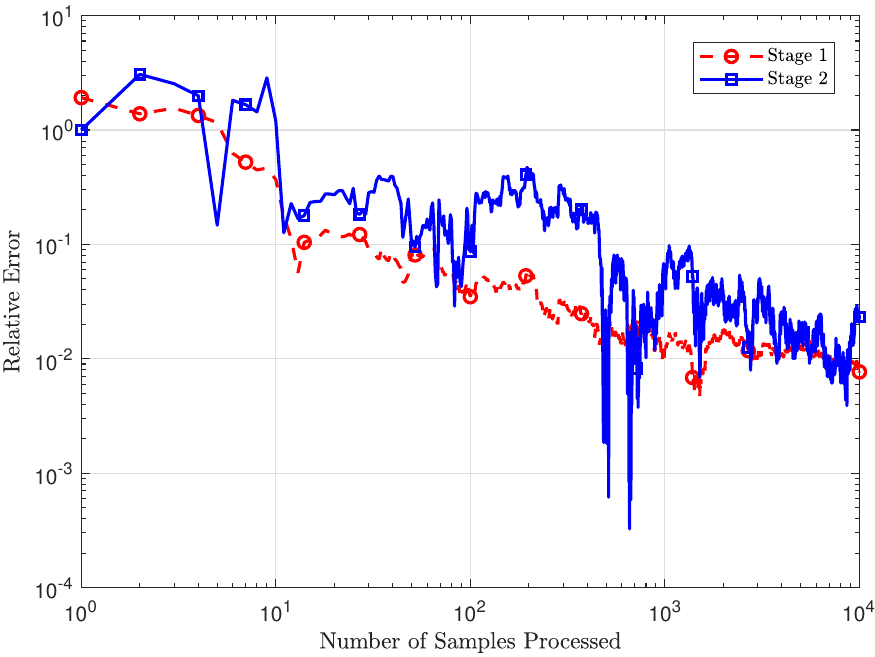}
	\caption{Convergence of the relative errors in Stage 1 (interpolation estimation) and Stage 2 (parameter identification).}
	\label{fig:stage1_error}
\end{figure}

The simulation results validate the effectiveness and robustness of the proposed two-stage algorithm. Figure~\ref{fig:stage1_error} depicts the convergence behavior of the interpolation estimation in Stage~1 and parameter reconstruction in Stage 2. To evaluate the overall accuracy of the identified parameters while avoiding bias from differences in their magnitudes, we adopt the Euclidean norm of the relative error vector as the performance metric:
\begin{equation}
	\label{eq_raletive_error}
	\begin{matrix}
		e_{\bar{\eta}}=\sqrt{\sum_i{\left( \frac{\widehat{\bar{\eta}}_i-\bar{\eta}_i}{\bar{\eta}_i} \right) ^2}},&		e_{\theta}=\sqrt{\left( \frac{\delta \theta _1}{\theta _1} \right) ^2+\left( \frac{\delta \theta _2}{\theta _2} \right) ^2}\\
	\end{matrix}
\end{equation}

As more steady-state samples are processed, the relative interpolation error $e_{\bar{\eta}}$ decreases steadily, indicating successful fusion of asynchronous local measurements. Subsequently, the relative parameter error $e_{\theta}$ also converges as samples proceed, demonstrating that the estimated interpolations provide a reliable basis for accurate structural parameter identification. Even under severe sampling irregularities, the proposed method achieves consistent and robust performance.

\begin{figure}[htbp]
	\centering
	\includegraphics[width=0.35\textwidth]{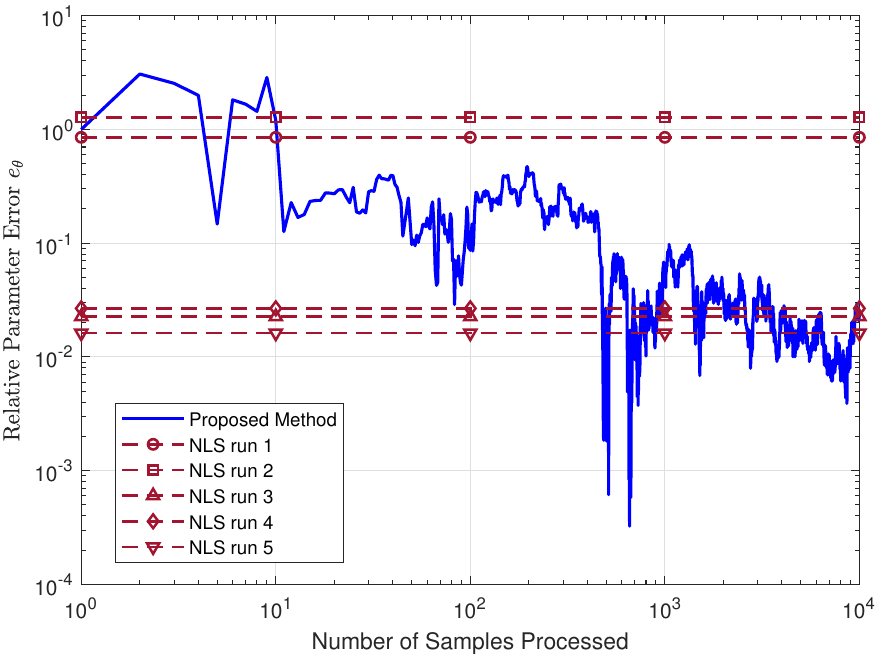}
	\caption{Relative error of the parameter estimation $\theta$ vs. the number of samples, compared with the final converged errors from 5 NLS runs.}
	\label{fig:param_error_comparison}
\end{figure}

For comparison, we also solved the identification problem using a NLS approach implemented via the MATLAB function \texttt{lsqnonlin}. In Figure \ref{fig:param_error_comparison}, the horizontal dashed lines represent the final converged error levels obtained from five independent NLS runs, each randomly initialized with a relative error of 10\% computed according to \eqref{eq_raletive_error}. The large variability among these runs reveals a well-known drawback of NLS: its tendency to become trapped in local minima, leading to inconsistent and often inaccurate estimates.

\begin{figure}[htbp]
	\centering
	\includegraphics[width=0.35\textwidth]{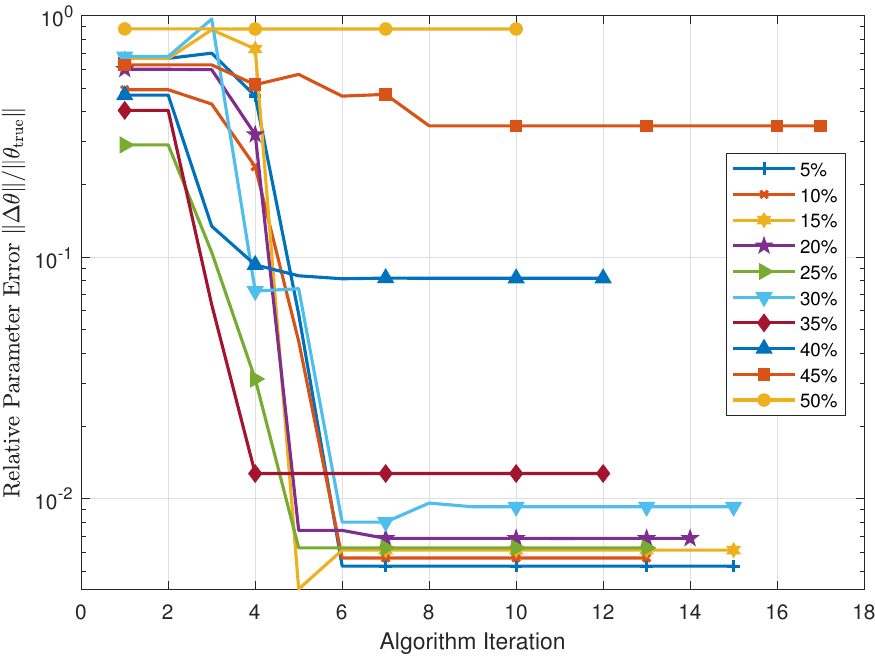}
	\caption{Convergence curves for NLS parameter estimation with 10 different random initializations.}
	\label{fig:nls_sensitivity}
\end{figure}

This sensitivity to initial conditions is further detailed in Figure \ref{fig:nls_sensitivity}, which shows the convergence paths for 10 separate NLS runs. Each run corresponds to a set of initial values with relative error magnitudes of 5\%, 10\%, …, up to 50\%. For each error magnitude, we generated 5 random initial values and selected the run with the smallest final error to illustrate how the relative error of the initial guess affects the ultimate parameter identification accuracy.

When the relative error between the initial guess and the true parameter values exceeds 45\%, the estimates produced by NLS become largely unreliable. This observation indicates that the NLS approach requires a reasonably accurate prior knowledge of the parameter values to achieve dependable results. In contrast, the proposed two-stage algorithm yields a stable and unique solution, effectively avoiding the pitfalls of local minima and the sensitivity to initialization. Moreover, for this large-scale system, the NLS method is computationally intensive, as it involves repeated simulations and relies on numerical gradient estimation due to the intractability of the analytical Jacobian.

\begin{remark}
	It was observed that the selection of interpolation points has a considerable impact on the identification accuracy, which may be related to the inherent sloppiness of the system \cite{Ma2025}. Furthermore, because the estimation process first involves solving for the matrix $X$ and subsequently recovering the parameters $\theta$, the method is sensitive to measurement noise and requires a relatively large amount of data to achieve reliable parameter estimates.
\end{remark}

\section{Conclusions}
\label{sec:conclusions}

This paper extend previous two-stage framework \cite{Zhou2024a} for identifying the interconnection structure of NDS under asynchronous, non-uniform, and slow-rate sampling. The first stage estimates right tangential interpolations from steady-state outputs, enabling the fusion of all asynchronous steady-state samples into a unified estimator. The second stage employs a left null-space projection to decouple the bilinear dependence between state-related matrices and interconnection parameters, leading to two linear estimation problems.

The method removes the full-normal-rank transfer matrix assumptions in prior work \cite{Zhou2024a}, extends applicability to MIMO NDS with arbitrary interconnections, and establishes theoretical guarantees of mean-square consistency and asymptotic unbiasedness. Numerical results confirm accurate structure recovery under severe sampling irregularities.

Future work will explore extensions using higher-order interpolations when repeated eigenvalues of the input generator are present, handling time-varying or nonlinear interconnections, and improving scalability for large-scale networks.



\appendix

\subsection{Proof of Theorem \ref{thm:steady_state_formula}}
\label{proof:thm_steady_state_formula}

This proof follows directly from the results of Lemma~\ref{lemma_output_decomposition} and Corollary~\ref{corollary_zero_order_moment}. From Lemma \ref{lemma_output_decomposition}, the steady-state response can be written as
\[
y_s(t) = \left(C_\theta X + D_{yu} \Pi\right) \xi(t).
\]

Corollary \ref{corollary_zero_order_moment} provides the explicit structure of the matrix term:
\begin{equation*}
	C_\theta X + D_{yu}\Pi = \begin{bmatrix}R&C\end{bmatrix}T^{-1}
\end{equation*}
where $R$ and $C$ are composed of the right tangential interpolation vectors corresponding to the real and complex eigenvalues of $\Xi$, respectively.

Let the Jordan decomposition of \(\Xi\) be \(\Xi = T\Lambda T^{-1}\). The input-generator state can thus be expressed as:
\begin{equation*}
	\xi(t) = e^{\Xi t}\xi(0) = T e^{\Lambda t} T^{-1} \xi(0)
\end{equation*}

Substituting this into the steady-state expression \( y_s(t) \) yields:
\begin{equation*}
	\begin{aligned}
		y_s(t) & =\left[ \begin{matrix}
			R & C \\
		\end{matrix} \right] T^{-1}Te^{\Lambda t}T^{-1}\xi (0) \\
		& =\left[ \begin{matrix}
			R & C \\
		\end{matrix} \right] e^{\Lambda t}T^{-1}\xi (0)        \\
	\end{aligned}
\end{equation*}

Using the block-diagonal structure of \(T\) and \(\Lambda\) from \eqref{eq_jordon}, together with the partitioned initial state \(\xi(0)\) from \eqref{eq_xi_0}, we can obtain:
\begin{equation*}
	\begin{aligned}
		T^{-1}\xi (0)&=\mathrm{col}\left\{ \left. \alpha _i \right|_{i=1}^{m_r},\left. \frac{1}{2}\left[ \begin{array}{c}
			\mu _i-j\nu _i\\
			\mu _i+j\nu _i\\
		\end{array} \right] \right|_{i=1}^{m_c} \right\}\\
		&=\mathrm{col}\left\{ \left. \alpha _i \right|_{i=1}^{m_r},\left. \frac{1}{2}\left[ \begin{array}{c}
			\beta _{i}^{*}\\
			\beta _i\\
		\end{array} \right] \right|_{i=1}^{m_c} \right\}\\
	\end{aligned}
\end{equation*}

The product \( \begin{bmatrix} R & C \end{bmatrix} e^{\Lambda t} \) is a block row vector whose components take the form
\begin{align*}
	\begin{bmatrix} R & C \end{bmatrix} e^{\Lambda t}
	&= \mathrm{row} \Big\{
	\left. e^{\lambda_{r,i} t} \eta_{r,i} \right|_{i=1}^{m_r}, \\
	&\quad \left. \begin{bmatrix}
		e^{\lambda_{c,i} t} \eta_{c,i} & e^{\lambda_{c,i}^* t} \eta_{c,i}^*
	\end{bmatrix} \right|_{i=1}^{m_c}
	\Big\}.
\end{align*}

Multiplication with the column vector \( T^{-1} \xi(0) \) then yields
\begin{equation*}
	\begin{aligned}
		y_s(t)= & \sum_{i=1}^{m_r}{\alpha _ie^{\lambda _{r,i}t}\eta _{r,i}}                                                                                  \\
		& +\sum_{i=1}^{m_c}{\frac{1}{2}\left( \beta _{i}^{*}e^{\lambda _{c,i}t}\eta _{c,i}+\beta _ie^{\lambda _{c,i}^{*}t}\eta _{c,i}^{*} \right)} \\
	\end{aligned}
\end{equation*}

Observing that the second term in each complex pair is the complex conjugate of the first, the sum becomes
\(
\frac{1}{2}\left( z + z^* \right) = \operatorname{Re}\{ z \},
\)
with \( z = \beta_i^{*} e^{\lambda_{c,i} t} \eta_{c,i} \).  

Therefore, the steady-state response simplifies to
\begin{equation*}
	y_s(t) = \sum_{i=1}^{m_r} \alpha_i e^{\lambda_{r,i}t}\eta_{r,i} + \sum_{i=1}^{m_c} \operatorname{Re}\{\beta_i^* e^{\lambda_{c,i}t}\eta_{c,i}\}
\end{equation*}
which completes the proof of Theorem \ref{thm:steady_state_formula}.

%
%

\subsection{Proof of Theorem \ref{thm:consistency}}
\label{proof:thm_consistency_appendix}
As described in Section \ref{sec:stage2}, the steady‑state output estimate $\widehat{Y}_{ss}$ is constructed by first reshaping the estimated right tangential interpolations $\widehat{\bar{\eta}}$ into the matrix $\begin{bmatrix}\widehat{R} & \widehat{C}\end{bmatrix}$, and then right‑multiplying by the constant matrix $T^{-1}$. This process makes it clear that $\widehat{Y}_{ss}$ is a linear transformation of the first‑stage estimate $\widehat{\bar{\eta}}$:
$$
\widehat{Y}_{ss} = \begin{bmatrix}\widehat{R} & \widehat{C}\end{bmatrix}T^{-1} = L_1\,\widehat{\bar{\eta}},
$$
where $L_1$ is a fixed, known matrix that implements the reshaping and right-multiplication by $T^{-1}$. Consequently, the estimation error satisfies
$$
\widehat{Y}_{ss}-Y_{ss} = L_1(\widehat{\bar{\eta}}-\bar{\eta}).
$$

Since Theorem \ref{thm:consistency_stage1} establishes the mean-square consistency of the first-stage estimator, namely $E\|\widehat{\bar{\eta}}-\bar{\eta}\|^2 \to 0$, it follows that $\widehat{Y}_{ss}$ likewise exhibits mean-square consistency:
\[
E\|\widehat{Y}_{ss}-Y_{ss}\|^2 \le \|L_1\|^2 E\|\widehat{\bar{\eta}}-\bar{\eta}\|^2 \to 0.
\]

Furthermore, since the first-stage least-squares is unbiased, $E[\widehat{\bar{\eta}}] = \bar{\eta}$, it follows that $E[\widehat{Y}_{ss}]=Y_{ss}$.

Proceeding to Step 2a, the vector $\gamma$ is an affine function of $\widehat{Y}_{ss}$, written as $\gamma = \gamma_0 + G\,\operatorname{vec}(\widehat{Y}_{ss})$. Under the identifiability condition, $\varGamma$ is a deterministic constant matrix with full column rank. The estimation error for $x = \operatorname{vec}(X)$ is therefore
$$
\widehat{x} - x = \varGamma^{\dagger}\Big(\gamma(\widehat{Y}_{ss})-\gamma(Y_{ss})\Big) = \varGamma^\dagger G\operatorname{vec}(\widehat{Y}_{ss}-Y_{ss}).
$$

Let $L_2=\varGamma ^{\dagger}G$, this linear relationship implies mean-square consistency for $\widehat{X}$:
$$
E\|\widehat{X}-X\|^2 \le \|L_2\|^2\,E\|\widehat{Y}_{ss}-Y_{ss}\|^2 \to 0.
$$
with unbiasedness also preserved, $E[\widehat{X}]=X$.

By Chebyshev's inequality, mean-square convergence implies convergence in probability, hence $\widehat{Y}_{ss}\xrightarrow{p} Y_{ss}$ and $\widehat{X}\xrightarrow{p} X$.

The least-squares solution for $\theta$ is given by
\[
\widehat{\theta} = \Psi(\widehat{X})^\dagger \kappa(\widehat{X},\widehat{Y}_{ss}).
\]

The estimation error $e_{\theta} := \widehat{\theta}-\theta$ satisfies the normal equation:
\[
\Psi(\widehat{X})^\top \Psi(\widehat{X})\,e_{\theta} = \Psi(\widehat{X})^\top r(\widehat{X},\widehat{Y}_{ss}),
\]
where $r(\widehat{X},\widehat{Y}_{ss}) := \kappa(\widehat{X},\widehat{Y}_{ss}) - \Psi(\widehat{X})\,\theta$.

Since at the true values $\Psi(X)\theta = \kappa(X,Y_{ss})$, the residual can be rewritten as:
$$
r(\widehat{X},\widehat{Y}_{ss}) = \big(\kappa(\widehat{X},\widehat{Y}_{ss})-\kappa(X,Y_{ss})\big) - \big(\Psi(\widehat{X})-\Psi(X)\big)\theta.
$$

The mappings $\Psi(\cdot)$ and $\kappa(\cdot,\cdot)$ are affine in their arguments and, together with the continuity of the Moore–Penrose pseudoinverse on sets of fixed rank, are Lipschitz continuous. Thus, there exist finite constants $L_{\Psi}, L_{\kappa,X}, L_{\kappa,Y}$ such that
\begin{align*}
	\|\Psi(\widehat{X})-\Psi(X)\|                             & \le L_{\Psi}\,\|\widehat{X}-X\|                                                 \\
	\|\kappa(\widehat{X},\widehat{Y}_{ss})-\kappa(X,Y_{ss})\| & \le L_{\kappa,X}\,\|\widehat{X}-X\| + L_{\kappa,Y}\,\|\widehat{Y}_{ss}-Y_{ss}\|
\end{align*}

This leads to a bound on the norm of the residual:
\begin{equation}
	\label{eq:residual_bound}
	\|r(\widehat{X},\widehat{Y}_{ss})\| \le \big(L_{\Psi}\|\theta\| + L_{\kappa,X}\big)\,\|\widehat{X}-X\| + L_{\kappa,Y}\,\|\widehat{Y}_{ss}-Y_{ss}\|
\end{equation}

Let $C_X=L_{\Psi}\|\theta\|+L_{\kappa, X}$ and $C_Y=L_{\kappa ,Y}$. From the normal equation, the error $e_\theta:=\widehat{\theta}-\theta$ is bounded by
\begin{equation} \label{eq:etheta_basic_bound}
	\|e_{\theta}\| \le \|\Psi(\widehat{X})^{\dagger}\|\,\|r(\widehat{X},\widehat{Y}_{ss})\| = \frac{\|r(\widehat{X},\widehat{Y}_{ss})\|}{\sigma_{\min}(\Psi(\widehat{X}))}
\end{equation}

By the identifiability condition, $\sigma_{\Psi} := \sigma_{\min}(\Psi(X)) > 0$ at the true parameters. Since singular values are continuous functions of the matrix entries, define event:
\[
\mathcal{G}\ :=\ \big\{\ \|\widehat{X}-X\|\ \le\ \delta_X\ \big\}
\]
where $\delta _X\,\,=\sigma _{\Psi}/2L_{\Psi}>0$.

On the event $\mathcal{G}$ one has
$$
\sigma_{\min}(\Psi(\widehat{X})) \ge \sigma_{\Psi} - L_{\Psi}\|\widehat{X}-X\| \ge \sigma_{\Psi}/2.
$$

Since $\widehat{X} \xrightarrow{p} X$, we have $\mathbb{P}(\mathcal{G}) \to 1$ as the number of samples $m_p \to \infty$. We can combine \refeq{eq:residual_bound} and \refeq{eq:etheta_basic_bound} to get:
$$
\|e_{\theta}\| \le \frac{2}{\sigma_{\Psi}}\Big(C_X\|\widehat{X}-X\| + C_Y\|\widehat{Y}_{ss}-Y_{ss}\|\Big).
$$

Since $\widehat{X} \to X$ and $\widehat{Y}_{ss} \to Y_{ss}$ in probability, the right-hand side of the inequality converges to 0 in probability, and hence $\widehat{\theta} \xrightarrow{p} \theta$.

\subsection{Proof of Theorem \ref{thm:ms_consistency}}
\label{proof:thm_ms_consistency_appendix}

We aim to show that $E\|e_{\theta}\|^2 \to 0$, where $e_{\theta} = \widehat{\theta}-\theta$. We split the expectation based on the event $\mathcal{G} = \{\|\widehat{X}-X\|\le \delta_X\}$ defined in the proof of Theorem \ref{thm:consistency}:
$$
E\|e_{\theta}\|^2 = E\big[\|e_{\theta}\|^2 \mathbf{1}_{\mathcal{G}}\big] + E\big[\|e_{\theta}\|^2 \mathbf{1}_{\mathcal{G}^c}\big]
$$
where $\mathbf{1}_{\mathcal{G}}$ is the indicator function for event $\mathcal{G}$, and $\mathcal{G}^c$ is its complement.

On the event $\mathcal{G}$, we can combine this with the error bound $\sigma _{\min}(\Psi (\widehat{X}))\ge \sigma _{\Psi}/2$ to get:
$$
\|e_{\theta}\| \le \frac{2}{\sigma_{\Psi}}\Big(C_X\|\widehat{X}-X\| + C_Y\|\widehat{Y}_{ss}-Y_{ss}\|\Big).
$$

Using the inequality $(a+b)^2 \le 2a^2+2b^2$, we can bound the expected squared error on $\mathcal{G}$:
\begin{align*}
	E\big[\|e_{\theta}\|^2 \mathbf{1}_{\mathcal{G}}\big] & \le \frac{8}{\sigma_{\Psi}^2}\Big(C_X^2 E\|\widehat{X}-X\|^2 + C_Y^2 E\|\widehat{Y}_{ss}-Y_{ss}\|^2\Big).
\end{align*}

From the proof of Theorem \ref{thm:consistency}, we know that $E\|\widehat{X}-X\|^2 \to 0$ and $E\|\widehat{Y}_{ss}-Y_{ss}\|^2 \to 0$. Therefore, the contribution from the event $\mathcal{G}$ vanishes as $m_p \to \infty$:
$$
E\big[\|e_{\theta}\|^2 \mathbf{1}_{\mathcal{G}}\big] \to 0.
$$

On the complement event $\mathcal{G}^c$, we use the general bound \refeq{eq:etheta_basic_bound} and apply the Cauchy-Schwarz inequality:
\begin{align*}
	E\big[\|e_{\theta}\|^2\,\mathbf{1}_{\mathcal{G}^c}\big] & = E\left[\frac{\|r(\cdot)\|^2}{\sigma_{\min}(\Psi(\widehat{X}))^2}\,\mathbf{1}_{\mathcal{G}^c}\right]                                       \\
	& \le \Big(E\big[\sigma_{\min}(\Psi(\widehat{X}))^{-4}\big]\Big)^{1/2} \Big(E\big[\|r(\cdot)\|^4\,\mathbf{1}_{\mathcal{G}^c}\big]\Big)^{1/2}
\end{align*}

Under the theorem's assumption, we have
$$
E\big[\|e_{\theta}\|^2\,\mathbf{1}_{\mathcal{G}^c}\big] \le \sqrt{C_{\Psi}} \cdot \Big(E\big[\|r(\cdot)\|^4\big]\Big)^{1/2}.
$$

Since $E\|r\|^4 \le 8(C_X^4 E\|\widehat{X}-X\|^4 + C_Y^4 E\|\widehat{Y}_{ss}-Y_{ss}\|^4)$ and standard results for LS imply the fourth moments of the errors in $\widehat{X}$ and $\widehat{Y}_{ss}$ also converge to zero, we have $E\|r\|^4 \to 0$.

Therefore, the contribution from the complement event also vanishes:
$$
E\big[\|e_{\theta}\|^2 \mathbf{1}_{\mathcal{G}^c}\big] \to 0.
$$

Combining both parts, we conclude that the total mean-square error converges to zero:
$$
E\|e_{\theta}\|^2 = E\big[\|e_{\theta}\|^2 \mathbf{1}_{\mathcal{G}}\big] + E\big[\|e_{\theta}\|^2 \mathbf{1}_{\mathcal{G}^c}\big] \to 0.
$$

This establishes mean-square consistency. Asymptotic unbiasedness follows directly from Jensen's inequality:
$$
\|E[\widehat{\theta}]-\theta\| = \|E[e_{\theta}]\| \le E[\|e_{\theta}\|] \le \left(E\|e_{\theta}\|^2\right)^{1/2} \to 0.
$$
Thus, $E[\widehat{\theta}] \to \theta$. This completes the proof.

\bibliographystyle{IEEEtran}
\bibliography{TCNS}

\end{document}